\documentclass{article}
\usepackage[T1]{fontenc}
\usepackage[utf8]{inputenc}
\usepackage{ismir}
\usepackage{amsmath,cite,url}
\usepackage{graphicx}
\usepackage{color}
\usepackage{multirow}

\usepackage[bookmarks=false,hidelinks]{hyperref}

\usepackage{textcomp}

\title{Universal Music Representations? Evaluating Foundation Models on World Music Corpora}

\multauthor
{Charilaos Papaioannou$^{1,2,3}$ \hspace{1cm} Emmanouil Benetos$^2$ \hspace{1cm} Alexandros Potamianos$^{1,3}$}{
$^1$ School of ECE, National Technical University of Athens, Greece\\
$^2$ Centre for Digital Music, Queen Mary University of London, UK\\
$^3$ Archimedes, Athena Research Center, Greece\\
{\tt\small cpapaioan@mail.ntua.gr}
}

\def\authorname{C. Papaioannou, E. Benetos, and A. Potamianos}

\begin{document}

\maketitle
\begin{abstract}
Foundation models have revolutionized music information retrieval, but questions remain about their ability to generalize across diverse musical traditions. This paper presents a comprehensive evaluation of five state-of-the-art audio foundation models across six musical corpora spanning Western popular, Greek, Turkish, and Indian classical traditions. We employ three complementary methodologies to investigate these models' cross-cultural capabilities: probing to assess inherent representations, targeted supervised fine-tuning of 1-2 layers, and multi-label few-shot learning for low-resource scenarios.  Our analysis shows varying cross-cultural generalization, with larger models typically outperforming on non-Western music, though results decline for culturally distant traditions. Notably, our approaches achieve state-of-the-art performance on five out of six evaluated datasets, demonstrating the effectiveness of foundation models for world music understanding. We also find that our targeted fine-tuning approach does not consistently outperform probing across all settings, suggesting foundation models already encode substantial musical knowledge. Our evaluation framework and benchmarking results contribute to understanding how far current models are from achieving universal music representations while establishing metrics for future progress.
\end{abstract}

\section{Introduction}\label{sec:intro}

The notion of music as a ``universal language'' remains contested among scholars \cite{mehr_universality_2019, savage_statistical_2015}. While some musical elements transcend cultural boundaries, traditions have evolved with distinct characteristics and semantic content \cite{trehub_cross-cultural_2015, margulis_narratives_2022}. This tension between universality and cultural specificity presents a complex challenge that modern artificial intelligence approaches offer a novel lens to investigate.

Foundation models have emerged as a transformative paradigm across artificial intelligence (AI) domains \cite{bommasani_opportunities_2022}, including music and audio \cite{li_mert_2024, wu_large-scale_2024, chu_qwen-audio_2023}. In music information retrieval (MIR), these multipurpose models perform diverse tasks from beat tracking to automatic tagging \cite{won_foundation_2023, ma_foundation_2024}. Though implicitly claiming a form of universality, they largely neglect cultural dimensions while training predominantly on Western-centric data \cite{ma_foundation_2024}. This raises a critical question: to what extent do foundation models actually provide universal music representations that generalize across diverse musical traditions?

In this work, we evaluate five state-of-the-art audio models across six corpora spanning Western popular, Greek, Turkish, and Indian classical traditions, to quantitatively assess their cross-cultural capabilities and contribute to discussions about the universality of musical representations. We focus on automatic music tagging as our evaluation task and employ three complementary methodologies: (i) probing, which uses the models as frozen feature extractors with a trainable classifier, (ii) targeted supervised fine-tuning to assess adaptation potential, and (iii) multi-label few-shot learning to evaluate performance in low-resource scenarios common with world music collections.

Our evaluation reveals both promising cross-cultural transfer capabilities as well as remaining gaps in universal music understanding, due to the decrease in performance for culturally distant domains and especially in low-resource scenarios. The contributions of this work can be summarized as follows:
\begin{itemize}
\item This is the first comprehensive evaluation, to the best of our knowledge, of foundation models across culturally diverse music corpora.
\item We propose a methodological evaluation framework that integrates few-shot learning with traditional approaches, enabling systematic assessment of model representations under different training setups.
\item State-of-the-art results have been achieved by our approaches in five out of six datasets.
\item We have optimized multi-label few-shot learning, significantly reducing inference time and making it practical for large numbers of classes.
\item Our code is being made available\footnote{\href{https://github.com/pxaris/FM-music-tagging}{\texttt{https://github.com/pxaris/FM-music-tagging}}} for reproducibility and to promote research on world music.
\end{itemize}

\section{Related Work}\label{sec:related}

\noindent\textbf{Foundation models.} Foundation models for music have emerged by leveraging large-scale self-supervised or contrastive learning on extensive audio datasets, enabling them to capture rich musical features applicable across diverse tasks. Representative works include JukeMIR \cite{castellon_codified_2021}, which explored representations from the Jukebox generative model \cite{dhariwal_jukebox_2020}, MULE \cite{mccallum_supervised_2022}, a self-supervised model pre-trained on MusicNet dataset, and Music2Vec \cite{li_map-music2vec_2022}, which utilized masked prediction strategies with student-teacher approaches. Subsequent advancements like MusicFM \cite{won_foundation_2023} have scaled up both model size and training data, demonstrating effectiveness across multiple benchmark tasks.

The landscape of current foundation models encompasses several architectural approaches: masked acoustic modeling, MERT \cite{li_mert_2024}, contrastive audio-text learning such as LAION-CLAP \cite{wu_large-scale_2024}, and unified audio understanding with models like Qwen-Audio \cite{chu_qwen-audio_2023}. Despite their impressive performance on standard benchmarks, their cross-cultural generalization capabilities remain largely unexplored, particularly regarding their effectiveness across diverse musical traditions beyond Western contexts.

\noindent\textbf{Automatic world music tagging.} Automatic music tagging - predicting metadata such as genre, mood, and instrumentation from audio signals - is typically referred to as music auto-tagging \cite{choi_deep_2018, kim_sample-level_2018, won_evaluation_2020, lee_sample-level_2017} and constitutes a multi-label classification problem. Architectures addressing this task have evolved from convolutional models like VGG-ish \cite{hershey_cnn_2017} and Musicnn \cite{pons_musicnn_2019} to transformer-based approaches like AST \cite{gong_ast_2021} and more recent foundation models \cite{won_foundation_2023}.

Research on world music computational analysis has grown in recent years \cite{panteli_computational_2018}, with studies focused on specific traditions including Turkish makam recognition \cite{demirel_automatic_2018, ganguli_critiquing_2022}, Indian classical music classification \cite{sharma_classification_2021}, and analysis of Iranian and Korean traditional music \cite{nikzat_kdc_2022,han_finding_2023}. While a recent study applied auto-tagging across diverse musical datasets \cite{papaioannou_west_2023}, this is the first time to the best of our knowledge where a comprehensive evaluation of foundation models on world music corpora is being conducted.

To address the challenges of imbalanced tags and limited data inherent in world music research, we employ Label-Combination Prototypical Networks (LC-Protonets) \cite{papaioannou_lc-protonets_2025} for few-shot learning. This approach extends Prototypical Networks \cite{snell_prototypical_2017} by creating prototypes for each label combination, rather than generating one prototype per label. While established benchmarks for evaluating representations on downstream tasks typically employ probing and fine-tuning methodologies \cite{turian_hear_2022, yang_superb_2021, yuan_marble_2023}, our work incorporates few-shot learning as a complementary evaluation approach, assessing foundation models' capabilities in low-resource scenarios.

\section{Methodological framework}\label{sec:method}

\begin{figure}[t]
    \centering
    \includegraphics[width=0.47\textwidth, alt={Flowchart showing three evaluation methodologies: Probing uses frozen foundation model with trainable MLP, Supervised Fine-Tuning unfreezes last layers plus MLP, and Multi-Label Few-Shot Learning extracts features from pretrained, probing, or fine-tuned models for prototype-based classification}]{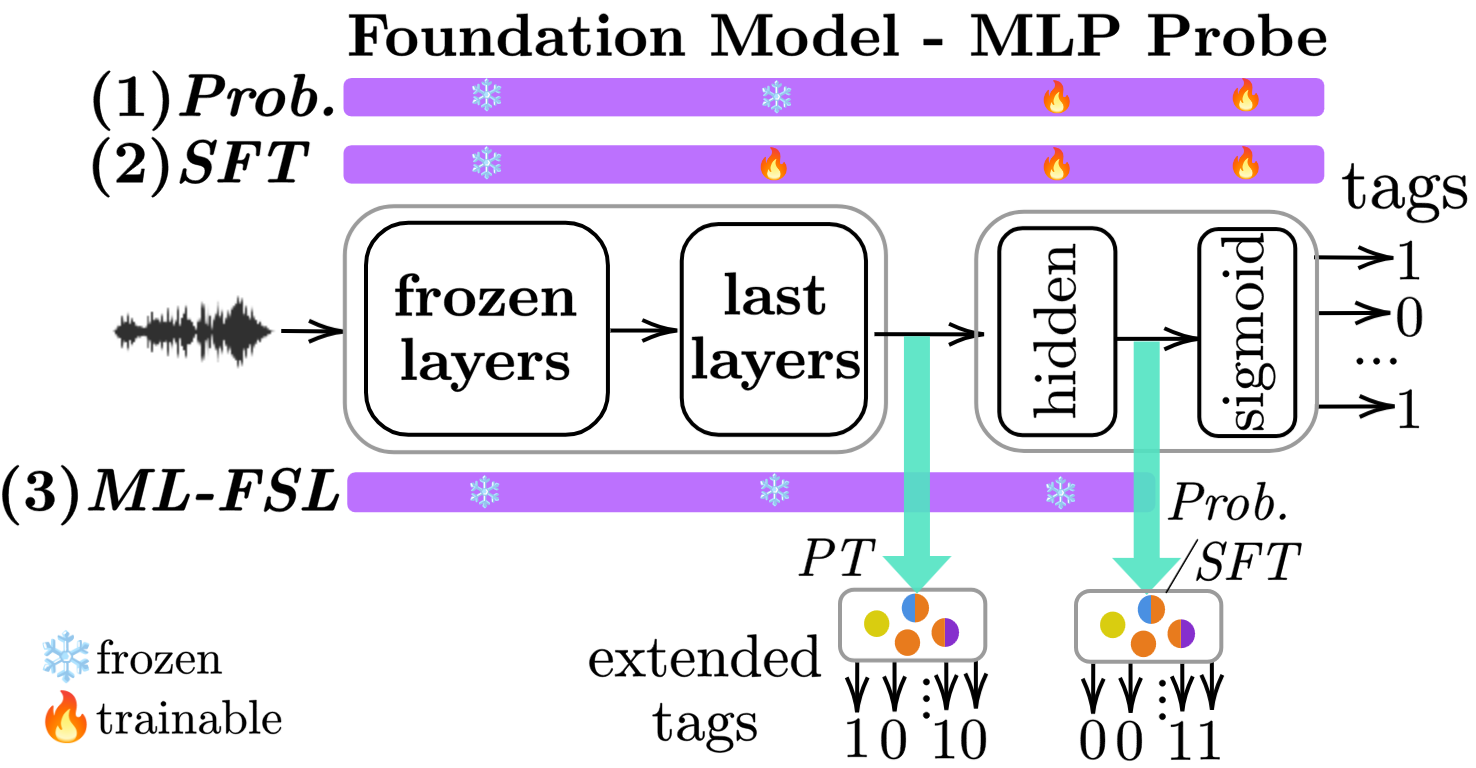}
    \caption{Architectural overview of our evaluation framework showcasing three methodologies: (1) Probing (\textit{Prob.}), (2) Supervised Fine-Tuning (\textit{SFT}), and (3) Multi-Label Few-Shot Learning (\textit{ML-FSL}). The diagram indicates feature extraction points used by \textit{ML-FSL} from either Pre-Trained (\textit{PT}), trained \textit{Prob.} or \textit{SFT} models.}
    \label{fig:architecture}
\end{figure}

Our methodological framework systematically evaluates whether foundation models can effectively represent musical characteristics across diverse cultural traditions. As shown in Figure~\ref{fig:architecture}, we employ three complementary methodologies: probing (\textit{Prob.}), supervised fine-tuning (\textit{SFT}), and multi-label few-shot learning (\textit{ML-FSL}). Probing trains only an MLP classifier on frozen model representations, while SFT makes the model's last layers trainable alongside the MLP. ML-FSL extracts representations from three contexts, i.e., pretrained model (\textit{PT}), trained probing model (\textit{Prob.}) and fine-tuned model (\textit{SFT}) to evaluate performance on extended tag sets under data scarcity conditions.

\subsection{Models}\label{subsec:models}
For our evaluation, we selected five state-of-the-art audio models spanning different architectures, pre-training approaches, and parameter scales:

\noindent\textbf{MERT.} We evaluate two variants of MERT \cite{li_mert_2024}: MERT-95M\footnote{\href{https://huggingface.co/m-a-p/MERT-v1-95M}{https://huggingface.co/m-a-p/MERT-v1-95M}} and MERT-330M\footnote{\href{https://huggingface.co/m-a-p/MERT-v1-330M}{https://huggingface.co/m-a-p/MERT-v1-330M}} with 95M and 330M parameters respectively. These transformer-based models employ masked acoustic modeling, using an acoustic and a musical teacher, during pre-training. MERT-95M consists of 12 layers, while MERT-330M has 24 layers.

\noindent\textbf{LAION-CLAP.} We include two variants: CLAP-Music\footnote{\href{https://huggingface.co/laion/larger_clap_music}{https://huggingface.co/laion/larger\_clap\_music}} (CLAP-M), trained exclusively on music data, and CLAP-Music\&Speech\footnote{\href{https://huggingface.co/laion/larger_clap_music_and_speech}{https://huggingface.co/laion/larger\_clap\_music\_and\_speech}} (CLAP-M\&S), which incorporates additional speech data \cite{wu_large-scale_2024}. Both utilize HTS-AT \cite{chen_hts-at_2022} for audio encoding, a transformer-based model with 4 groups of swin-transformer blocks \cite{liu_swin_2021}, with 68M audio-specific parameters within a larger 194M parameter model.

\noindent\textbf{Qwen2-Audio.} The largest model in our evaluation framework, Qwen2-Audio\footnote{\href{https://huggingface.co/Qwen/Qwen2-Audio-7B}{https://huggingface.co/Qwen/Qwen2-Audio-7B}} \cite{chu_qwen2-audio_2024}, contains 637M audio-specific parameters within an 8.4B parameter architecture and features 32 transformer layers \cite{vaswani_attention_2017} in its audio tower.

\noindent\textbf{VGG-ish.} As a baseline comparison, we include VGG-ish \cite{simonyan_very_2015, won_evaluation_2020}, a 3.6M parameter end-to-end model trained via supervised learning on mel-spectrograms to predict tags. For VGG-ish, we report results from the literature for the same experimental setup used in our work \cite{papaioannou_west_2023, papaioannou_lc-protonets_2025} rather than running new experiments.

\subsection{Datasets}\label{subsec:datasets}

Our evaluation spans diverse traditions from six music datasets. For Western music, we utilize MagnaTagATune \cite{law_evaluation_2009} (25,863 clips) and FMA-medium \cite{defferrard_fma_2017} (25,000 tracks). For world music traditions, we incorporate the Lyra dataset \cite{papaioannou_dataset_2022} with 1,570 recordings of Greek folk music, and three collections from the CompMusic project \cite{serra_creating_2014}: the Turkish-makam corpus \cite{uyar_corpus_2014, senturk_computational_2016} (5,297 recordings) as well as Hindustani \cite{srinivasamurthy_corpora_2014} (1,204 recordings) and Carnatic \cite{srinivasamurthy_corpora_2014} (2,612 recordings) of Indian classical music.

Following \cite{papaioannou_west_2023}, we set maximum audio durations to achieve similar sizes between datasets and prepare their metadata for the auto-tagging task. For Probing and Supervised Fine-Tuning, we use the standard tag sets, i.e., $50$ tags for MagnaTagATune, $30$ for Lyra and Turkish-makam, and $20$ for the rest of the datasets. Our ML-FSL experiments use extended tag sets that include previously unseen classes, summing up to: $80$ tags for MagnaTagATune, $60$ for Lyra and Turkish-makam, $40$ for FMA-medium and Carnatic, and $35$ for Hindustani, consistent with \cite{papaioannou_lc-protonets_2025}.

\subsection{Evaluation methodologies}\label{subsec:methodologies}

\noindent\textbf{Probing.} Our first methodology (\textit{Prob.}) evaluates how well foundation models inherently represent musical characteristics across cultures. We employ probing, where the model remains frozen while only training a classifier on top of the extracted representations. Specifically, we implement a shallow Multi-layer Perceptron (MLP) with a single hidden layer of $512$ units followed by a sigmoid classification layer, optimized with binary cross-entropy loss.

\noindent\textbf{Supervised Fine-Tuning.} To evaluate adaptation potential, we implement targeted supervised fine-tuning (\textit{SFT}) by unfreezing a subset of model parameters. For MERT-95M, we unfreeze the last two transformer layers, while for MERT-330M only the last layer. For both CLAP models, we unfreeze the last group of swin-transformer blocks of the audio encoder along with the normalization and two projection layers. In Qwen2-Audio, we fine-tune the last layer of the audio tower along with the normalization layer before multi-modal projection. These choices were constrained by RAM limitations affecting both trainable parameters and hyperparameter tuning. We use the same trainable MLP Probe architecture as in the Probing experiments, initializing it with the weights learned during that phase. This weight initialization strategy helps maintain previously learned knowledge while adapting to new domains, mitigating potential catastrophic forgetting issues \cite{kirkpatrick_overcoming_2017}. We also employ learning rate warmup and cosine scheduling to ensure stable adaptation \cite{gupta_continual_2023}.

\noindent\textbf{Multi-Label Few-Shot Learning.} Our third methodology (\textit{ML-FSL}) evaluates performance in low-resource scenarios by employing an optimized version of LC-Protonets \cite{papaioannou_lc-protonets_2025} that is detailed in subsection \ref{subsec:mlfsl_optim}. We extract representations from three different contexts: directly from the pre-trained model (\textit{PT}), from the hidden layer of the trained MLP Probe (\textit{Prob.}), and from the fine-tuned model (\textit{SFT}). Notably, this methodology involves no additional training during few-shot evaluation; the model acts as a frozen feature extractor that maps both the few examples and the unknown items to an embedding space where classification occurs utilizing the LC-Protonets approach.

\begin{figure}[t]
    \centering
    \includegraphics[width=0.42\textwidth,  alt={Scatter plot showing positive correlation between model size (x-axis, logarithmic scale from 10^0 to 10^3 audio parameters in millions) and ROC-AUC performance (y-axis, 84-89 percent), with five models plotted: VGG-ish, CLAP-Music\&Speech, MERT variants, and Qwen2-Audio}]{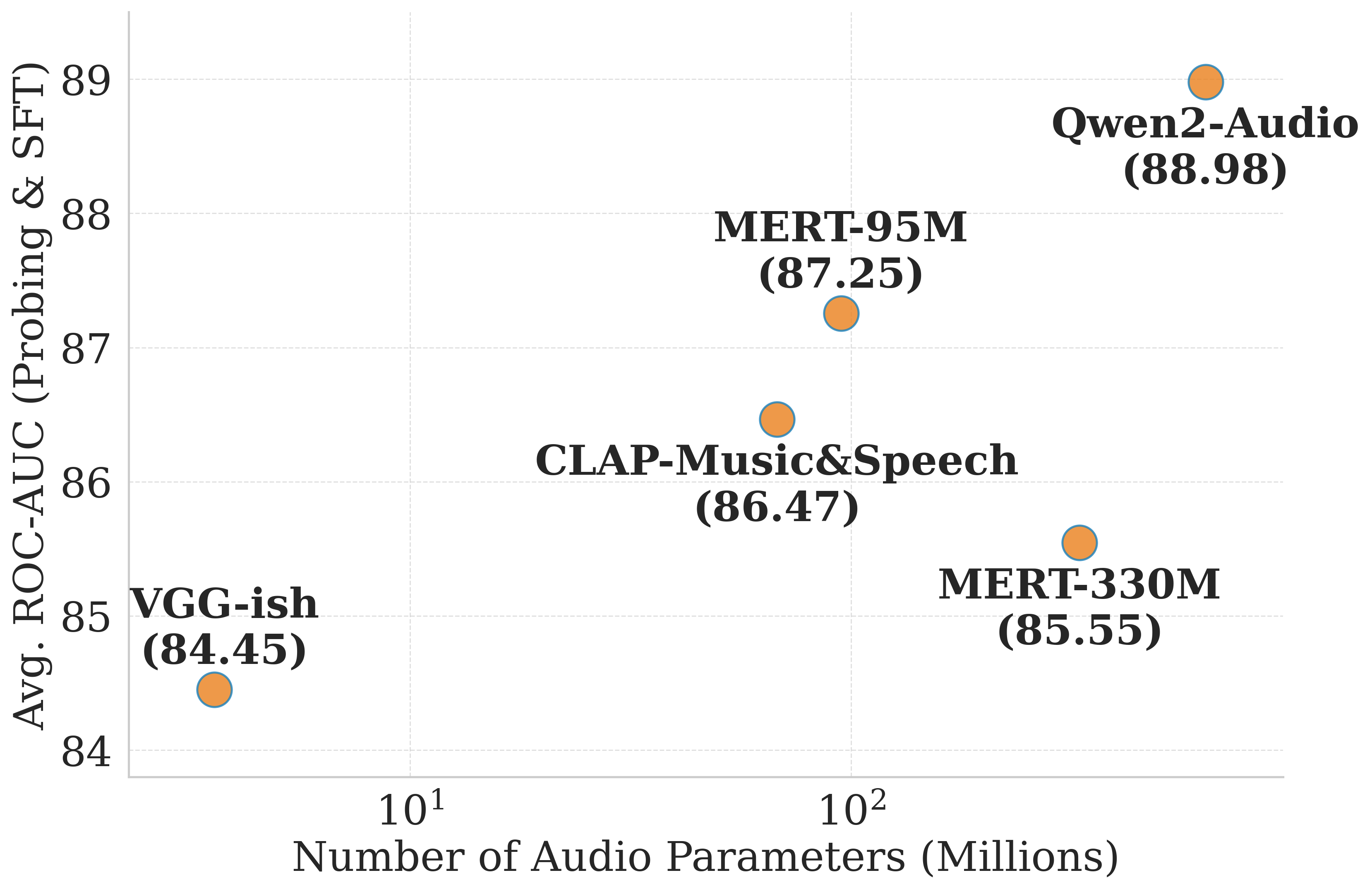}
    \caption{Relationship between model size and performance, averaged over Probing and Supervised Fine-Tuning (SFT) tasks. The x-axis represents the number of audio-specific parameters on a logarithmic scale, while the y-axis reports the mean ROC-AUC (\%) across all datasets.}
    \label{fig:probing_sft}
\end{figure}

\subsection{Multi-label few-shot learning optimization}\label{subsec:mlfsl_optim}

While the LC-Protonets method \cite{papaioannou_lc-protonets_2025} offers significant performance advantages for multi-label few-shot learning, its computational complexity increases substantially with the number of labels due to the exponential growth of label combinations. In this work, we introduce an optimization that significantly improves inference efficiency while maintaining identical classification results.

The original approach creates an LC-Prototype (LCP) for each label combination (LC-class) derived from the power sets of the few available examples' labels. Each available example is called a \textit{support item} and it is defined by $(\mathbf{x}_i, \mathbf{y}_i)$, with $\mathbf{x}_i$ being its input feature vector and $\mathbf{y}_i$ the set of its labels. For the set of support items $S$, the set of all LC-classes $L$ is computed as $L = \bigcup_{(\mathbf{x}_i, \mathbf{y}_i) \in S}^{} \mathcal{P}(\mathbf{y}_i)$, where $\mathcal{P}(\mathbf{y_i})$ is the power set of the labels of the $i$-th support item, excluding the empty set. For each LC-class $L_j$, with $j = 1,2,...,|L|$, the LCP representation $\mathbf{p}_j$ is computed by averaging the embeddings of all support items that include $L_j$ in their power sets:
\begin{equation}
\mathbf{p}_j = \frac{1}{|S_j|}\sum_{(\mathbf{x}_i, \mathbf{y}_i) \in S_j}^{}f_\theta(\mathbf{x}_i),
\end{equation}
where $S_j = \{(\mathbf{x}_i, \mathbf{y}_i) \in S \mid L_j \in \mathcal{P}(\mathbf{y_i})\}$, and $f_\theta$ the embedding mapping model.

Our key insight is that multiple LC-classes often share identical LCP representations despite representing different label combinations. This occurs because the same set of support items contributes to multiple label combinations derived from their power sets. For example, if a support item with labels $\{A, B, C\}$ is the only item contributing to both $\{A, B\}$ and $\{B, C\}$ LCPs, these LCPs will have identical representations.

We exploit this redundancy by maintaining a dictionary structure that maps unique LCP representations to their corresponding sets of LC-classes:
\begin{equation}
\text{UniqueLCPs} = \{\mathbf{p}_m \mapsto \{L_j \mid \mathbf{p}_j = \mathbf{p}_m\}\},
\end{equation}
where $j = 1,2,...,|L|$ and $m = 1,2,...,M$ with $M$ being the number of unique LCPs and $M \ll |L|$.
During inference, instead of computing distances between a \textit{query item}, unseen during training, and all possible $|L|$ LCPs, we only compute distances to the $M$ unique LCP representations. For the nearest unique LCP, we then select the label combination with the maximum cardinality, consistent with the original LC-Protonets method.

Our experiments show that this approach yields speed improvements of 10× for datasets with $20$ labels, scaling to more than 100× for datasets with $60$ labels, while producing identical classification results to the original method. We apply this optimization to the LC-Protonets repository\footnote{\href{https://github.com/pxaris/LC-Protonets}{\texttt{https://github.com/pxaris/LC-Protonets}}}, making it practical for large label sets.

\section{Experimental setup}\label{sec:experiments}

\noindent\textbf{Experiments and resources.} We conducted 5 runs with different random seeds for both Probing and ML-FSL tasks, but a single run for SFT due to computational constraints. SFT trainable parameters varied: 14M for MERT-95M, 13M for MERT-330M, 25M for CLAP models, and 56M for Qwen2-Audio. All experiments ran on an NVIDIA RTX A5000 GPU, and we used Qwen2-Audio in half-precision (FP16) in all our methodologies to fit in this card. Most SFT training completed within 24 hours, with only 3 out of 30 experiments extending to about 36 hours.

\noindent\textbf{Dataset processing.} We standardized Turkish-makam, Hindustani, and Carnatic datasets to approximately 200 hours each, matching MagnaTagATune and FMA-medium durations \cite{papaioannou_west_2023}, while Lyra remained at its original 80 hours. We followed the training, validation, and test splits from \cite{won_evaluation_2020, papaioannou_west_2023}. For ML-FSL, evaluation items came exclusively from test sets \cite{papaioannou_lc-protonets_2025} to prevent data leakage. 

\noindent\textbf{Model-specific configurations.} Each foundation model required specific preprocessing: MERT models use 30-second windows at 24kHz, CLAP models 10-second windows at 48kHz, and Qwen2-Audio 30-second windows at 16kHz. All audio was converted to mono and resampled to the model's required rate. 

\noindent\textbf{Representation extraction strategies.} For MERT models, we extract representations by summing the average, across time, hidden states of the last four layers of the models. For CLAP models, we extract them from the audio projection layer which takes as input the average pooled layer representation of the last hidden state. For Qwen2-Audio, we use the last hidden state embeddings averaged across all layers of the whole model, when passing a simple text prompt that includes nothing but the respective tags for audio processing, i.e., \small\texttt{<|audio\_bos|><|AUDIO|><|audio\_eos|>}\normalsize. These representation extraction strategies, number of fine-tuned layers, and other design choices of our method were optimized through preliminary experiments.

\noindent\textbf{Hyperparameters.} For Probing, we used Adam optimizer \cite{kingma_adam_2017} ($\beta_1=0.9$, $\beta_2=0.999$, $\epsilon=10^{-8}$) with learning rate $10^{-3}$, batch size 16, early stopping patience 10, and maximum 200 epochs. For SFT, we used AdamW \cite{loshchilov_decoupled_2019} with identical $\beta$ parameters but learning rate $10^{-4}$, model-specific batch sizes (to fit maximum available resources) with gradient accumulation to simulate batch size 16 across all setups, patience 5, and maximum 30 epochs. We applied learning rate warmup and cosine scheduling for the first 5\% of SFT epochs. ML-FSL evaluations used cosine distance with an $N$-way $K$-shot setup, with $N$ being the number of extended tags per dataset and $K$ equal to $3$ examples per label in all experiments. We also attempted Low-Rank Adaptation \cite{hu_lora_2021} initially but abandoned it due to extensive hyperparameter tuning requirements across our $5\times6$ experimental matrix.

\noindent\textbf{Evaluation metrics.} For the Probing and SFT methodologies, we report area under the receiver operating characteristic curve (ROC-AUC) and mean average precision (mAP). These metrics are particularly well-suited for multi-label classification tasks \cite{davis_relationship_2006} and are consistent with prior work in music tagging \cite{won_evaluation_2020, papaioannou_west_2023}. For ML-FSL evaluation, we report macro-F1 (M-F1) and micro-F1 (m-F1) scores, which align with the LC-Protonets evaluation framework \cite{papaioannou_lc-protonets_2025}. F1 score is the harmonic mean of the precision and recall scores. Macro-F1 gives equal weight to all classes, while micro-F1 accounts for class imbalance by calculating metrics globally across all instances.

\begin{table}[t]
    \centering
    \renewcommand{\arraystretch}{1.1}
    \setlength\tabcolsep{4.4pt}
    \resizebox{0.48\textwidth}{!}{%
    \begin{tabular}{l|c|cc|cc}
    \hline
    \multirow{2}{*}{\textbf{Model}} & \textbf{Params} & \multicolumn{2}{c|}{\multirow{2}{*}{\textbf{ROC-AUC} (\%)}} & \multicolumn{2}{c}{\multirow{2}{*}{\textbf{mAP} (\%)}} \\
    & Audio/Total & \multicolumn{2}{c|}{} & \multicolumn{2}{c}{} \\
    \hline
    VGG-ish \cite{papaioannou_west_2023} & \small{$3.6$M/$3.6$M} & \multicolumn{2}{c|}{84.45} & \multicolumn{2}{c}{50.56} \\
    \hline
    & & \underline{\textit{Prob.}} & \underline{\textit{SFT}} & \underline{\textit{Prob.}} & \underline{\textit{SFT}} \\
    \textbf{MERT-95M} & \small{$95$M/$95$M} & 87.25$_{0.32}$ & 87.26 & 52.25$_{0.42}$ & 52.68 \\
\textbf{MERT-330M} & \small{$330$M/$330$M} & 85.40$_{0.68}$ & 85.69 & 49.62$_{0.83}$ & 50.47 \\
\textbf{CLAP-M} & \small{$68$M/$194$M} & 71.52$_{1.14}$ & 78.96 & 29.98$_{1.07}$ & 40.41 \\
\textbf{CLAP-M\&S} & \small{$68$M/$194$M} & 86.78$_{0.31}$ & 86.15 & 53.12$_{0.87}$ & 51.99 \\
\textbf{Qwen2-Audio} & \small{$637$M/$8.40$B} & \textbf{88.59$_{0.47}$} & \textbf{89.37} & \textbf{56.48$_{0.63}$} & \textbf{58.73} \\
\hline
    \end{tabular}}
    \caption{Model performance comparison averaged across all datasets for Probing and SFT tasks. Values are averaged over multiple runs with subscripted standard deviations. Bold values indicate best performance per column.}
    \label{tab:probing_sft_results}
    \end{table}

\begin{table*}[ht]
    \centering
    \renewcommand{\arraystretch}{1.0}
    \setlength\tabcolsep{2.8pt}
    \resizebox{\textwidth}{!}{%
    \begin{tabular}{l|cc|cc|cc|cc|cc|cc}
    \hline
    \multirow{2}{*}{\textbf{Model}} & \multicolumn{2}{c|}{\textbf{MagnaTagATune}} & \multicolumn{2}{c|}{\textbf{FMA-medium}} & \multicolumn{2}{c|}{\textbf{Lyra}} & \multicolumn{2}{c|}{\textbf{Turkish-makam}} & \multicolumn{2}{c|}{\textbf{Hindustani}} & \multicolumn{2}{c}{\textbf{Carnatic}} \\
    & ROC-AUC & mAP & ROC-AUC & mAP & ROC-AUC & mAP & ROC-AUC & mAP & ROC-AUC & mAP & ROC-AUC & mAP \\
    \hline
    VGG-ish \cite{papaioannou_west_2023} & 91.23 & 45.82 & 88.89 & 49.49 & 80.97 & 48.06 & 86.96 & 56.39 & 84.77 & 60.82 & 73.92 & 42.78 \\
    \hline
    \multicolumn{13}{c}{\textit{Probing (Prob.)}} \\
    \hline
    \textbf{MERT-95M} & 90.46$_{0.10}$ & 44.16$_{0.21}$ & 91.68$_{0.08}$ & 51.43$_{0.43}$ & 85.61$_{0.66}$ & 53.34$_{0.61}$ & \textbf{88.22$_{0.23}$} & 57.89$_{0.34}$ & 86.59$_{0.52}$ & 60.26$_{0.56}$ & 80.96$_{0.35}$ & 46.41$_{0.35}$ \\
\textbf{MERT-330M} & 89.66$_{0.16}$ & 41.73$_{0.59}$ & 90.78$_{0.11}$ & 48.85$_{0.32}$ & 84.65$_{0.78}$ & 51.81$_{0.59}$ & 85.37$_{0.64}$ & 52.45$_{1.12}$ & 84.23$_{1.36}$ & 58.78$_{2.08}$ & 77.73$_{1.03}$ & 44.07$_{0.31}$ \\
\textbf{CLAP-M} & 80.07$_{0.21}$ & 25.82$_{0.13}$ & 77.42$_{0.15}$ & 22.89$_{0.38}$ & 64.18$_{1.29}$ & 31.16$_{0.43}$ & 77.31$_{0.51}$ & 38.77$_{1.00}$ & 68.69$_{4.05}$ & 33.43$_{4.21}$ & 61.47$_{0.60}$ & 27.83$_{0.30}$ \\
\textbf{CLAP-M\&S} & 92.41$_{0.05}$ & \textbf{48.54$_{0.16}$} & 94.05$_{0.08}$ & 59.13$_{0.54}$ & 87.25$_{0.18}$ & 56.94$_{0.51}$ & 86.49$_{0.27}$ & 54.69$_{0.36}$ & 82.61$_{1.14}$ & 55.70$_{3.29}$ & 77.85$_{0.13}$ & 43.73$_{0.35}$ \\
\textbf{Qwen2-Audio} & 91.17$_{0.13}$ & 45.58$_{0.21}$ & 96.60$_{0.07}$ & 73.38$_{0.28}$ & 86.44$_{0.81}$ & 53.50$_{0.65}$ & 86.64$_{0.42}$ & 53.38$_{0.79}$ & \textbf{88.45$_{0.83}$} & 62.42$_{0.99}$ & 82.22$_{0.56}$ & 50.59$_{0.88}$ \\
\hline
    \multicolumn{13}{c}{\textit{Supervised Fine-Tuning (SFT)}} \\
    \hline
    \textbf{MERT-95M} & 90.62 & 44.52 & 91.70 & 51.74 & 84.89 & 53.62 & 87.50 & \textbf{57.91} & 88.20 & 61.47 & 80.64 & 46.83 \\
\textbf{MERT-330M} & 89.55 & 41.93 & 91.12 & 49.56 & 84.74 & 52.54 & 86.17 & 53.80 & 85.49 & 61.33 & 77.05 & 43.66 \\
\textbf{CLAP-M} & 88.54 & 39.26 & 88.37 & 42.04 & 71.97 & 38.14 & 79.82 & 42.49 & 75.65 & 45.01 & 69.39 & 35.51 \\
\textbf{CLAP-M\&S} & 91.77 & 47.54 & 92.86 & 57.11 & 85.35 & 52.86 & 86.69 & 54.93 & 83.73 & 56.91 & 76.51 & 42.58 \\
\textbf{Qwen2-Audio} & 92.03 & 48.27 & \textbf{97.02} & \textbf{75.94} & \textbf{87.57} & \textbf{57.04} & 87.95 & 56.10 & 88.32 & \textbf{64.35} & \textbf{83.35} & \textbf{50.66} \\
\hline
(Previous) SOTA & \textbf{92.7} & 46.54 & 92.4 & 53.7 & 85.4 & 54.3 & 87.7 & 57.7 & 86.5 & 63.1 & 77.0 & 43.9 \\
\hline
    \end{tabular}}
    \caption{Model performance on individual datasets for Probing and SFT tasks. For Probing, values are averaged over multiple runs with subscripted standard deviations, while SFT results are from single runs. Bold values indicate best performance per metric and dataset. SOTA values are from \cite{huang_mulan_2022} for MagnaTagATune and \cite{papaioannou_west_2023} for the rest of the datasets.}
    \label{tab:probing_sft_datasets}
    \end{table*}

\section{Results}\label{sec:results}
    
\subsection{Probing and Supervised Fine-Tuning}\label{subsec:probing_sft}

Table \ref{tab:probing_sft_results} presents the performance of the evaluated foundation models averaged across all datasets for both Probing and SFT tasks. Overall, Qwen2-Audio achieves the highest performance with 88.59\% ROC-AUC and 56.48\% mAP in Probing, further improving to 89.37\% ROC-AUC and 58.73\% mAP after fine-tuning. This is followed by MERT-95M and CLAP-Music\&Speech with comparable performance, while CLAP-Music shows significantly lower performance without speech data in its training corpus. 

Figure \ref{fig:probing_sft} illustrates the relationship between model size (audio-specific parameters) and ROC-AUC performance, averaged across datasets and both Probing and SFT tasks. A generally positive correlation is revealed, with similar trends observed in both methodologies. Qwen2-Audio (637M parameters) consistently outperforms smaller models, achieving 88.98\% average ROC-AUC score. Surprisingly, MERT-95M (87.25\%) outperforms the much larger MERT-330M (85.55\%). This is worth noting as \cite{yuan_marble_2023} reported that both models performed on par for auto-tagging tasks, suggesting that our common representation extraction strategy for both MERT models may not optimally leverage the larger model's capacity. Another potential explanation is that MERT-95M has been trained on open data whereas MERT-330M has been trained with additional proprietary data with a strong Western bias \cite{li_mert_2024}.

When examining Probing (\textit{Prob.}) performance across individual datasets, in Table \ref{tab:probing_sft_datasets}, we observe a consistent pattern of decreasing performance for music traditions that are culturally distant from the data used to pre-train the respective foundation models. Western music datasets (MagnaTagATune and FMA-medium) consistently achieve the highest performance across all models, with ROC-AUC values reaching 96.60\% for Qwen2-Audio on FMA-medium. Greek (Lyra) and Turkish (makam) music datasets show moderate performance, while Indian classical music (Hindustani and Carnatic) datasets typically exhibit the lowest performance. This cultural performance gap is especially pronounced for CLAP-Music, where the ROC-AUC drops from 80.07\% for MagnaTagATune to 61.47\% for Carnatic.

Applying Supervised Fine-Tuning (\textit{SFT}) generally improves performance across all models and datasets, with an average gain of 1-2\% in ROC-AUC for most models. Notably, CLAP-Music shows the largest improvement with \textit{SFT}, indicating greater adaptation potential despite lower absolute performance. For other models, the modest gains suggest that they require broader fine-tuning to further shift their pre-trained representations towards different cultures.

Importantly, our approaches achieve state-of-the-art performance in five out of six datasets, with MagnaTagATune being the only exception. However, their consistent performance decrease towards diverse cultures, suggests that their representations are still biased toward Western musical traditions.

\subsection{Multi-label few-shot learning}\label{subsec:mlfsl}

Table \ref{tab:mlfsl_results} presents the ML-FSL evaluation results averaged across all datasets using extended tag sets. The results show consistent performance improvements moving from pre-trained models (PT) to trained probing models (Prob.) and then to supervised fine-tuned models (SFT) across all foundation models. The substantial gap between macro-F1 and micro-F1 metrics indicates considerable class imbalance in the extended tag sets, while the increased standard deviation stems from the support set sampling which can significantly impact the classification performance.

\begin{table}[t]
    \centering
    \renewcommand{\arraystretch}{1.1}
    \setlength\tabcolsep{1.4pt}
    \resizebox{0.48\textwidth}{!}{%
    \begin{tabular}{l|ccc|ccc}
    \hline
    \textbf{Model} & \multicolumn{3}{c|}{\textbf{M-F1}} & \multicolumn{3}{c}{\textbf{m-F1}} \\
    \hline
    VGG-ish \cite{papaioannou_lc-protonets_2025} & \multicolumn{3}{c|}{30.18} & \multicolumn{3}{c}{55.09} \\
    \hline
    & \underline{\textit{PT}} & \underline{\textit{Prob.}} & \underline{\textit{SFT}} & \underline{\textit{PT}} & \underline{\textit{Prob.}} & \underline{\textit{SFT}} \\
    \textbf{MERT-95M} & 23.90$_{1.52}$ & 28.05$_{1.74}$ & 28.28$_{1.80}$ & 46.59$_{1.57}$ & 52.16$_{1.43}$ & 52.56$_{1.63}$ \\
\textbf{MERT-330M} & 23.03$_{1.12}$ & 28.48$_{1.40}$ & 28.51$_{1.28}$ & 45.11$_{1.29}$ & 51.78$_{1.51}$ & 51.80$_{1.46}$ \\
\textbf{CLAP-M} & 17.71$_{1.20}$ & 18.43$_{1.40}$ & 21.58$_{1.13}$ & 38.80$_{1.37}$ & 39.97$_{1.20}$ & 46.57$_{1.20}$ \\
\textbf{CLAP-M\&S} & \textbf{28.23$_{1.36}$} & 29.22$_{1.09}$ & 30.27$_{1.90}$ & \textbf{51.59$_{1.54}$} & 53.32$_{1.31}$ & 54.43$_{1.27}$ \\
\textbf{Qwen2-Audio} & 25.98$_{1.36}$ & \textbf{30.96$_{1.26}$} & \textbf{32.00$_{1.41}$} & 49.97$_{1.41}$ & \textbf{55.66$_{0.82}$} & \textbf{56.85$_{1.23}$} \\
\hline
    \end{tabular}}
    \caption{ML-FSL performance averaged across datasets on extended tag sets. Results show macro-F1 (M-F1) and micro-F1 (m-F1) across contexts (\textit{PT}, \textit{Prob.}, \textit{SFT}). Values are means with subscripted standard deviations. Bold indicates best performance per column.}
    \label{tab:mlfsl_results}
    \end{table}

\begin{table*}[ht]
    \centering
    \renewcommand{\arraystretch}{1.0}
    \setlength\tabcolsep{4.4pt}
    \resizebox{\textwidth}{!}{%
    \begin{tabular}{l|cc|cc|cc|cc|cc|cc}
    \hline
    \multirow{2}{*}{\textbf{Model}} & \multicolumn{2}{c|}{\textbf{MagnaTagATune}} & \multicolumn{2}{c|}{\textbf{FMA-medium}} & \multicolumn{2}{c|}{\textbf{Lyra}} & \multicolumn{2}{c|}{\textbf{Turkish-makam}} & \multicolumn{2}{c|}{\textbf{Hindustani}} & \multicolumn{2}{c}{\textbf{Carnatic}} \\
    & M-F1 & m-F1 & M-F1 & m-F1 & M-F1 & m-F1 & M-F1 & m-F1 & M-F1 & m-F1 & M-F1 & m-F1 \\
    \hline
    VGG-ish \cite{papaioannou_lc-protonets_2025} & 26.40 & 37.31 & 29.12 & 45.37 & 46.05 & 69.03 & \textbf{30.07} & \textbf{56.22} & 31.33 & 58.38 & 18.13 & \textbf{64.25} \\
    \hline\multicolumn{13}{c}{\textit{Pre-Trained models (PT)}} \\
    \hline
    \textbf{MERT-95M} & 18.76$_{1.04}$ & 28.37$_{1.38}$ & 16.24$_{0.64}$ & 35.37$_{0.94}$ & 46.87$_{2.59}$ & 66.07$_{2.25}$ & 20.69$_{1.77}$ & 40.95$_{1.80}$ & 25.87$_{2.45}$ & 51.50$_{1.92}$ & 14.97$_{0.64}$ & 57.26$_{1.10}$ \\
\textbf{MERT-330M} & 18.17$_{0.78}$ & 26.99$_{1.36}$ & 16.24$_{0.69}$ & 31.15$_{1.51}$ & 44.22$_{1.45}$ & 65.48$_{1.57}$ & 20.14$_{2.01}$ & 39.71$_{1.95}$ & 25.08$_{1.40}$ & 50.14$_{1.08}$ & 14.32$_{0.41}$ & 57.21$_{0.28}$ \\
\textbf{CLAP-M} & 13.10$_{0.84}$ & 20.00$_{1.15}$ & 9.65$_{0.29}$ & 19.77$_{1.31}$ & 33.56$_{2.88}$ & 57.14$_{1.49}$ & 14.33$_{1.10}$ & 32.12$_{1.37}$ & 21.06$_{1.63}$ & 47.38$_{1.60}$ & 14.55$_{0.43}$ & 56.42$_{1.30}$ \\
\textbf{CLAP-M\&S} & 25.90$_{0.55}$ & 36.55$_{0.61}$ & 28.78$_{1.66}$ & 42.95$_{2.02}$ & 48.03$_{2.02}$ & 69.04$_{1.54}$ & 24.19$_{1.73}$ & 47.13$_{2.22}$ & 26.29$_{1.20}$ & 54.50$_{1.57}$ & 16.19$_{1.02}$ & 59.38$_{1.30}$ \\
\textbf{Qwen2-Audio} & 21.29$_{0.51}$ & 32.09$_{0.26}$ & 29.76$_{2.23}$ & 47.50$_{1.86}$ & 39.99$_{1.05}$ & 64.24$_{1.07}$ & 19.89$_{1.71}$ & 42.27$_{1.88}$ & 28.42$_{1.96}$ & 55.92$_{1.70}$ & 16.55$_{0.69}$ & 57.82$_{1.67}$ \\
\hline
    \multicolumn{13}{c}{\textit{Trained Probing models (Prob.)}} \\
    \hline
    \textbf{MERT-95M} & 23.77$_{0.85}$ & 34.71$_{1.03}$ & 24.62$_{1.19}$ & 42.96$_{1.30}$ & 45.80$_{2.76}$ & 68.16$_{1.81}$ & 26.14$_{1.73}$ & 50.00$_{0.70}$ & 30.75$_{2.95}$ & 56.41$_{2.18}$ & 17.25$_{0.98}$ & 60.70$_{1.55}$ \\
\textbf{MERT-330M} & 24.48$_{0.59}$ & 34.78$_{1.45}$ & 25.21$_{0.76}$ & 40.65$_{1.76}$ & 47.92$_{3.26}$ & \textbf{70.15$_{2.18}$} & 26.97$_{1.61}$ & 50.47$_{1.13}$ & 29.25$_{1.55}$ & 53.77$_{1.82}$ & 17.06$_{0.61}$ & 60.85$_{0.69}$ \\
\textbf{CLAP-M} & 14.84$_{0.49}$ & 22.67$_{1.00}$ & 11.55$_{0.50}$ & 22.72$_{1.48}$ & 34.85$_{4.03}$ & 57.73$_{1.37}$ & 16.68$_{0.81}$ & 36.00$_{1.22}$ & 18.77$_{1.42}$ & 44.96$_{0.95}$ & 13.87$_{1.16}$ & 55.74$_{1.15}$ \\
\textbf{CLAP-M\&S} & 26.90$_{0.47}$ & 37.62$_{0.93}$ & 31.14$_{1.28}$ & 46.53$_{1.59}$ & 47.10$_{0.89}$ & 69.77$_{0.53}$ & 25.58$_{1.59}$ & 49.70$_{1.39}$ & 28.11$_{1.38}$ & 56.43$_{2.19}$ & 16.46$_{0.92}$ & 59.88$_{1.25}$ \\
\textbf{Qwen2-Audio} & 26.79$_{0.40}$ & 37.65$_{0.21}$ & 39.49$_{1.02}$ & 56.30$_{0.82}$ & 42.52$_{1.81}$ & 67.10$_{1.13}$ & 26.09$_{1.65}$ & 51.59$_{1.20}$ & 31.62$_{1.26}$ & 60.08$_{0.40}$ & 19.25$_{1.40}$ & 61.24$_{1.14}$ \\
\hline
    \multicolumn{13}{c}{\textit{Supervised Fine-Tuned models (SFT)}} \\
    \hline
    \textbf{MERT-95M} & 24.46$_{0.79}$ & 35.28$_{0.90}$ & 24.94$_{1.18}$ & 42.78$_{1.44}$ & 45.51$_{3.74}$ & 67.93$_{2.72}$ & 26.16$_{1.87}$ & 49.76$_{1.54}$ & 30.40$_{2.15}$ & 56.39$_{1.68}$ & 18.18$_{1.08}$ & 63.19$_{1.48}$ \\
\textbf{MERT-330M} & 23.78$_{0.65}$ & 33.67$_{0.91}$ & 24.94$_{1.21}$ & 39.95$_{1.77}$ & \textbf{48.50$_{2.75}$} & 70.06$_{2.23}$ & 26.84$_{1.51}$ & 50.29$_{1.25}$ & 30.56$_{1.31}$ & 55.25$_{1.58}$ & 16.43$_{0.27}$ & 61.57$_{1.04}$ \\
\textbf{CLAP-M} & 22.15$_{0.51}$ & 32.67$_{1.22}$ & 19.61$_{0.79}$ & 34.81$_{0.99}$ & 30.46$_{2.04}$ & 55.86$_{2.02}$ & 20.66$_{1.69}$ & 45.80$_{1.13}$ & 21.95$_{1.31}$ & 50.74$_{1.14}$ & 14.63$_{0.45}$ & 59.53$_{0.67}$ \\
\textbf{CLAP-M\&S} & 26.28$_{0.50}$ & 37.23$_{1.09}$ & 30.27$_{1.56}$ & 46.57$_{1.61}$ & 48.09$_{4.74}$ & 69.93$_{2.28}$ & 28.91$_{1.75}$ & 53.87$_{1.56}$ & 31.27$_{2.47}$ & 57.41$_{0.74}$ & 16.82$_{0.37}$ & 61.55$_{0.34}$ \\
\textbf{Qwen2-Audio} & \textbf{27.67$_{0.25}$} & \textbf{38.57$_{0.18}$} & \textbf{40.10$_{1.29}$} & \textbf{57.17$_{0.95}$} & 44.13$_{2.45}$ & 68.34$_{2.38}$ & 27.61$_{2.37}$ & 53.98$_{1.55}$ & \textbf{32.52$_{1.23}$} & \textbf{60.26$_{0.89}$} & \textbf{19.97$_{0.87}$} & 62.76$_{1.43}$ \\
\hline
    \end{tabular}}
    \caption{ML-FSL performance on extended tag sets per dataset. Results show macro-F1 (M-F1) and micro-F1 (m-F1) across three contexts. Values are means with subscripted standard deviations. Bold indicates best performance per column.}
    \label{tab:mlfsl_combined_datasets}
    \end{table*}

Qwen2-Audio demonstrates the best overall performance in the ML-FSL task with 32.00\% macro-F1 and 56.85\% micro-F1 after fine-tuning, followed closely by CLAP-Music\&Speech with 30.27\% macro-F1 and 54.43\% micro-F1. Notably, even the best foundation model's performance (Qwen2-Audio) is comparable to a VGG-ish feature extractor trained via supervised learning on standard tags for each dataset. This stands in contrast to the Probing and SFT settings (Table \ref{tab:probing_sft_results}), where foundation models clearly outperform VGG-ish, showing that ML-FSL tasks remain challenging for them despite their extensive pre-training. Supervised learning of a VGG-ish model on extended tag sets has not been conducted in the literature, likely due to the scarcity of examples for infrequent tags.

\begin{figure}[t]
    \centering
    \includegraphics[width=0.4\textwidth, alt={Line graph comparing original versus optimized LC-Protonets method, showing number of labels (20-60) on x-axis, with two y-axes: number of LC-Prototypes (left, logarithmic scale 10^0 to 10^5) and inference time per item (right, logarithmic scale 10^0 to 10^5 milliseconds). Original method shows exponential growth while optimized method remains nearly flat}]{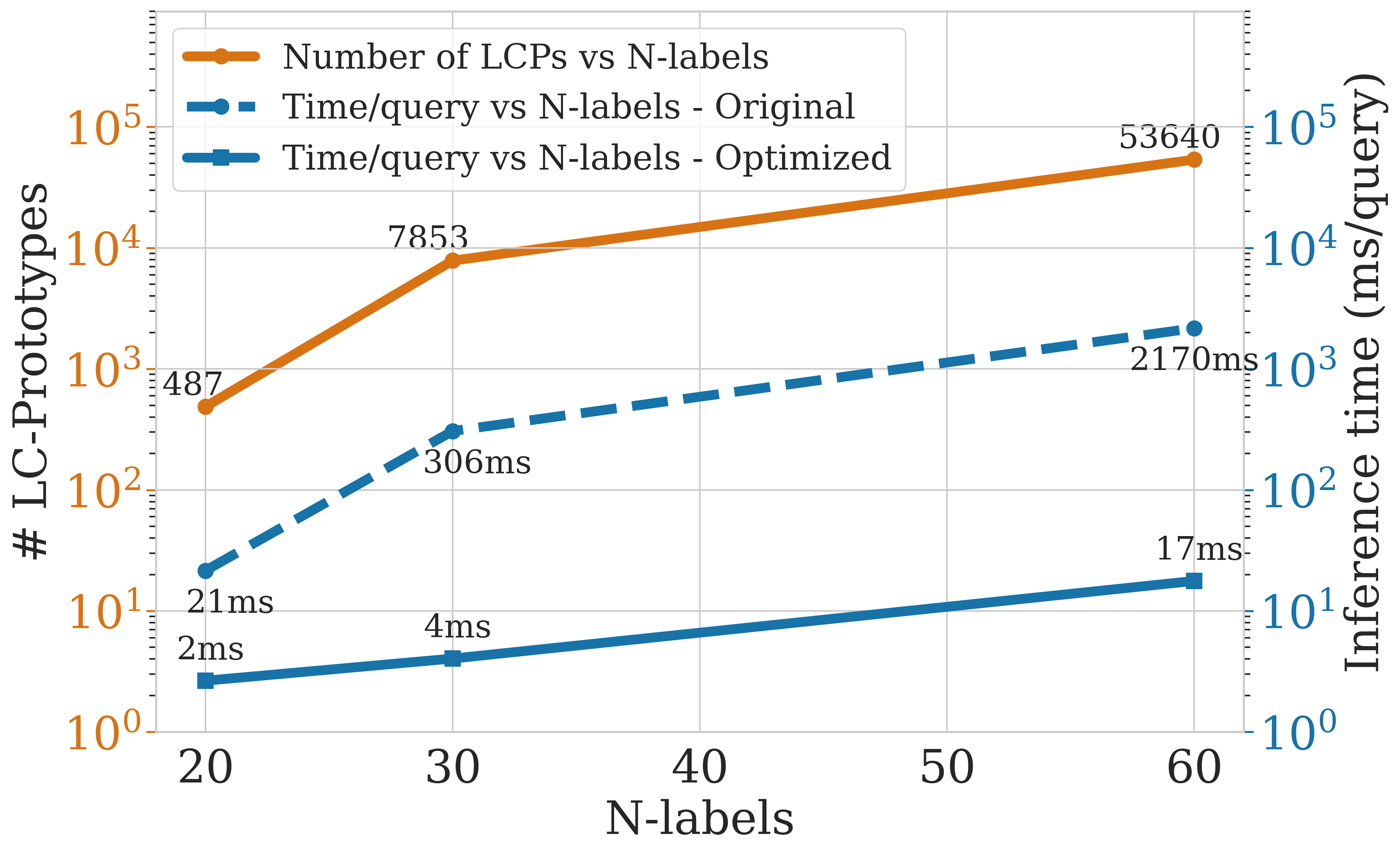}
    \caption{Scalability metrics of the LC-Protonets method, averaged across all datasets. The $x$-axis represents the number of labels, the left $y$-axis shows the number of LCPs, and the right $y$-axis indicates the inference time per item with both $y$-axes using the same logarithmic scale.}
    \label{fig:lcp_optimization}
\end{figure}

When examining the ML-FSL results per dataset in Table \ref{tab:mlfsl_combined_datasets}, we observe that only on Western datasets (MagnaTagATune and FMA-medium) does the best foundation model (Qwen2-Audio) achieve significantly better performance than the VGG-ish baseline. For Turkish-makam, VGG-ish representations actually outperform foundation models, while for Lyra, Hindustani, and Carnatic, the results are comparable. This pattern provides additional clear evidence of the implicit Western-centric bias integrated into models due to their pre-training data.

\noindent\textbf{LC-Protonets optimization.}
Figure \ref{fig:lcp_optimization} illustrates the scalability metrics of our optimized LC-Protonets approach compared to the original method, averaged across datasets. As the number of labels increases from $20$ to $60$, the number of LC-Prototypes grows exponentially, from approximately 500 to over 50,000. This growth leads the original method to a corresponding increase from 21ms to over 2,000ms inference time per query item (dashed blue line). However, our optimization (solid blue line), leveraging the unique prototypes, mitigates the computational complexity issues, requiring only 2ms in the $20$ labels cases and rising to no more than 20ms for $60$ labels, a 100× improvement.

\section{Conclusions}\label{sec:conclusions}

In this paper, we examined the universality of music representations in foundation models through a comprehensive methodological framework evaluating five state-of-the-art audio models across six world music corpora. Although these models achieved better performance than previous models for diverse music traditions, we found clear indicators of Western-centric bias.

Our incorporation of ML-FSL tasks particularly revealed this limitation. When faced with these challenging scenarios, foundation models performed on par with significantly smaller and simpler models, with performance notably degrading further on non-Western datasets. 

To further enable ML-FSL evaluation, we substantially optimized the computational complexity of the utilized method, by forming unique prototypes representing multiple label combinations. We demonstrated that this change makes it practical for large sets of labels, a typical condition when studying world music datasets.

Future work could extend our methodological framework by incorporating Low-Rank Adaptation (LoRA) and implement broader supervised fine-tuning to investigate further cultural adaptation. More tasks can also be included such as mode estimation, exploring the analogies between key on Western cultures and makam or raga recognition in other cultures.  

We hope this work brings attention to the cultural dimensions of foundation models while providing a framework for quantitatively assessing progress toward truly universal musical representations.

\pagebreak

\section{Acknowledgments}\label{sec:acknowledgments}

We would like to thank the reviewers for their valuable and constructive feedback, which helped us improve our study. This work has been partially supported by project MIS 5154714 of the National Recovery and Resilience Plan Greece 2.0 funded by the European Union under the NextGenerationEU Program.

\bibliography{bibliography}

\end{document}